\documentclass[prd,showpacs,aps,eqsecnum,nofootinbib,twocolumn]{revtex4}

\usepackage{amsmath,amsfonts,amssymb}

\begin{document}

\title{Dynamics of test bodies with spin in de Sitter spacetime}

\author{Yuri N.\ Obukhov}
\email{obukhov@math.ucl.ac.uk}
\affiliation{Department of Mathematics and Institute of Origins, 
University College London, Gower Street, London, WC1E 6BT, United Kingdom}

\author{Dirk Puetzfeld}
\email{dirk.puetzfeld@aei.mpg.de}
\homepage{http://www.aei.mpg.de/~dpuetz/}
\affiliation{Max-Planck-Institute for Gravitational Physics 
(Albert-Einstein-Institute), Am Muehlenberg 1, 14476 Golm, Germany}

\begin{abstract}
We study the motion of spinning test bodies in the de Sitter spacetime of constant positive curvature. With the help of the 10 Killing vectors, we derive the 4-momentum and the tensor of spin explicitly in terms of the spacetime coordinates. However, in order to find the actual trajectories, one needs to impose the so-called supplementary condition. We discuss the dynamics of spinning test bodies for the cases of the Frenkel and Tulczyjew conditions.
\end{abstract}



\maketitle

\section{Introduction}

The study of motion of test bodies in General Relativity (GR) theory has a long history, see for example the account in \cite{Havas:1989}. Among other approaches, the multipole approximation method represents a powerful technique with the help of which one can derive a self-consistent set of equations of motion for a body characterized by the moments of arbitrary order. In the zeroth -- or pole -- order one recovers the geodesic equation in the context of multipolar methods. At the next -- first or pole-dipole -- order, the test body is described by the 4-momentum and the tensor of spin, and the dynamics is governed by the Mathisson-Papapetrou equations, the relevant discussion can be found in \cite{Mathisson:1937,Shanmugadhasan:1947,Papapetrou:1951:3,Pirani:1956,Tulczyjew:1959,Tulczyjew:1962,Taub:1964,Dixon:1964,Beiglboeck:1965,Dixon:1965:1,Madore:1969,Dixon:1970:1,Dixon:1970:2,Dixon:1973:1,Dixon:1974:1,Dixon:1979,Puetzfeld:Obukhov:2007,Dixon:2008:1,Steinhoff:Puetzfeld:2010}. Even in the absence of spin the integration of the equations of motion is a difficult problem for nontrivial spacetimes. The spinless test body moves along a timelike geodesic on the curved manifold. When the spin is non-zero, the motion becomes nontrivial even in the flat spacetime \cite{Frenkel:1926,Weyssenhoff:Raabe:1947}. Moreover, in curved spacetime the motion is no longer geodesic due to the Lorentz-like force that acts on the test body. The Mathisson-Papapetrou force depends on the spacetime curvature and this considerably complicates the problem of finding the trajectories, \cite{Corinaldesi:Papapetrou:1951,Mashhoon:1971,Plyatsko:1998,Mohseni:2002,Semerak:1999,Kyrian:Semerak:2007}. 

Here we investigate the dynamics of test bodies with spin in de Sitter spacetime. The latter is the maximally symmetric 4-dimensional space which means that there exist 10 ($=4\times(4+1)/2$) Killing vector fields that describe the symmetries of this manifold. The flat Minkowski spacetime also has 10 Killing vectors, and in this respect the geometrical properties of the de Sitter spacetime are close to those of the Minkowski space. At the same time, the de Sitter manifold has a nontrivial curvature. 

We should clearly stress at this point, that this work is {\it not} concerned with the actual derivation of the equations of motion in the context of different multipolar approximation schemes. We will not discuss the conceptual questions which eventually lead to the imposition of different supplementary conditions, nor do we discuss different ``flavors'' of multipolar schemes -- in particular not the subtleties regarding the definition of the moments within these schemes. Here we are only concerned with the solution of the pole-dipole equations motions in a specific background spacetime for two frequently used supplementary conditions \cite{Frenkel:1926,Pirani:1956,Tulczyjew:1959}. As our analysis will show, different choices of the supplementary condition lead to quite different dynamics. In other words, the selection of such a condition -- which in the context of the multipolar schemes under consideration has the status of {\it additional} assumption on the level of the equations of motion -- should be performed with utmost care. These assumptions should not be misunderstood as the initial conditions. The number of equations of motion is less than the number of dynamical variables (see the next section), and the supplementary condition makes the system self-consistent and predictive.

We should also stress that our analysis is valid for both interpretations of the Mathisson-Papapetrou equations which can be found in the literature, i.e.\ it applies to point particles \cite{Corben:1968,Wong:1972,Kannenberg:1977,Audretsch:1981} as well as to extended test bodies. Recall that -- on the level of the equations motion -- the descriptions of both types of objects formally coincide. One should keep in mind though, that there are preferences regarding the supplementary condition \cite{Beiglboeck:1967,Schattner:1978,Schattner:1979}, depending on the system which is supposed to be described by the equations of motion.  

The structure of the paper is as follows. In section \ref{eom_sec} we present a short overview of the equations of motion for spinning test bodies. We pay particular attention to the conserved quantities in these equations. In section \ref{de_sitter_space_sec} we collect some facts about the Killing vectors of de Sitter spacetime. These results are then used for the integration of the equations of motion in de Sitter spacetime in section \ref{eom_integration_sec}. This is followed by a discussion of how momentum and spin can be expressed with the help of the integrals motion in section \ref{integrals_of_motion_sec}. Finally, we draw our conclusion in \ref{conclusions_sec} and present a brief outlook on open problems. A summary of our conventions and a directory of symbols can be found in the appendix \ref{app_sec}.   

\section{Equations of motion}\label{eom_sec}

A test body with spin is described by the following variables. Its trajectory is given by 4 spacetime coordinates $x^\alpha(s)$ as functions of the affine parameter $s$ (proper time). Furthermore, the body is characterized by the first two moments (pole and dipole) coming from its energy-momentum contents: the 4-momentum $p^\alpha$ and the spin $S^{\alpha\beta} = -\,S^{\beta\alpha}$. One can interpret these as the two ``gravitational charges'' carried by the body since they couple to the gravitational field in the same way as the electric charge couples to the electromagnetic field. 

The pole-dipole equations of motion (usually known as the Mathisson-Papapetrou equations) read
\begin{eqnarray}\label{dP}
\dot{p}^\alpha &=& -\,{\frac 12}\,S^{\mu\nu}u^\beta\,R_{\mu\nu\beta}{}^\alpha,\\
\dot{S}^{\alpha\beta}  &=&  2p^{[\alpha}\,u^{\beta]}.\label{dS}
\end{eqnarray}
Here $u^\alpha = dx^\alpha/ds$ is the 4-velocity of the body, and the dot denotes the covariant derivative with respect to the proper time, $``\dot{\phantom{aa}}"= D/ds = u^\alpha\nabla_\alpha$. The force term on the right-hand side of (\ref{dP}) depends explicitly on the spacetime curvature. 

The set of the Mathisson-Papapetrou equations is insufficient to determine the dynamics of the system. Indeed, we have 14 unknown variables ($x^\alpha, p^\alpha, S^{\mu\nu}$) and only 10 equations (\ref{dP})-(\ref{dS}). One additional algebraic equation comes in the form of the normalization condition $u^\alpha u_\alpha = 1$.\footnote{Since the dynamics of a spinning particle is non-geodesic, in general, this cannot be viewed as an integral of motion, but is a mere constraint due to the choice of parametrization of the trajectories.} Thus one needs 3 more equations to make the model predictable. Such additional equations are usually algebraic and they are commonly known as the supplementary conditions. Although this name is rather misleading, we will keep the tradition. The two most widely used are the Frenkel condition \cite{Frenkel:1926} (sometimes also called Pirani condition \cite{Pirani:1956})
\begin{equation}
S^{\alpha\beta}u_\beta = 0, \quad (\ast) \label{FC}
\end{equation}
and the Tulczyjew condition \cite{Tulczyjew:1959}
\begin{equation}
S^{\alpha\beta}p_\beta = 0. \quad (\ast\ast) \label{TC}
\end{equation}
Both algebraic equations have 3 independent components and thus the total number of the equations becomes equal to the number of the unknowns. The Frenkel condition (\ref{FC}) was introduced in the electrodynamical context by taking into account the purely ``magnetic-dipole'' nature of electron's spin, and it is nowadays mainly used in a point particle context. Although also Tulczyjew's condition (\ref{TC}) was initially used in a point-particle representation, it is nowadays more often associated with extended bodies.

In this paper we will analyze the dynamics of spinning test bodies for both supplementary conditions. 

By contracting (\ref{dS}) with $u_\beta$, we find
\begin{equation}
p^\alpha = mu^\alpha + \dot{S}^{\alpha\beta}u_\beta.\label{pmu}
\end{equation}
The $m := u^\alpha p_\alpha$ we will call the rest mass of the body, defined as usual as the projection of the 4-momentum on the rest frame of moving body. Besides that, we can define another mass parameter by $M^2 := p^\alpha p_\alpha$. In general, these two masses are different. We will compare them below. Depending on the supplementary condition chosen, the mass parameters may be constant or not. 

\subsection{Conserved quantity}

Let $\xi$ be a Killing vector. This is a solution of the equation $\nabla_\alpha \xi_\beta + \nabla_\beta\xi_\alpha = 0$. Applying $\nabla_\gamma$, we derive
\begin{equation}
\nabla_\gamma\nabla_\alpha\xi_\beta + \nabla_\gamma\nabla_\beta\xi_\alpha = \nabla_\gamma\nabla_\alpha\xi_\beta + \nabla_\beta\nabla_\gamma\xi_\alpha - R_{\gamma\beta\alpha}{}^\lambda\xi_\lambda = 0.
\end{equation}
Now, add and subtract $\nabla_\alpha\nabla_\beta\xi_\gamma$ and use the identity $\nabla_\gamma\nabla_\alpha\xi_\beta + \nabla_\beta\nabla_\gamma\xi_\alpha + \nabla_\alpha\nabla_\beta\xi_\gamma \equiv 0$. As a result, we obtain the second covariant
derivative of any Killing vector in terms of the curvature:
\begin{equation}\label{ddx}
\nabla_\alpha\nabla_\beta\xi_\gamma = R_{\beta\gamma\alpha}{}^\lambda\xi_\lambda.
\end{equation}
Then we straightforwardly find
\begin{eqnarray}
{\frac D {ds}}\left(2\xi_\alpha p^\alpha\right) &=& 2\dot{\xi}_\alpha p^\alpha + 2\xi_\alpha\dot{p}^\alpha,\label{dPx}\\
{\frac D {ds}}\left(S^{\alpha\beta}\nabla_\alpha\xi_\beta\right) &=& -\,2\dot{\xi}_\alpha p^\alpha + S^{\alpha\beta}u^\gamma R_{\alpha\beta\gamma}{}^\lambda\xi_\lambda.\label{dSdx}
\end{eqnarray}
In the last equation we used (\ref{dS}) and (\ref{ddx}). Taking the sum of (\ref{dPx}) and (\ref{dSdx}), and using the equation of motion (\ref{dP}), we derive
\begin{equation}
{\frac D {ds}}\left(2\xi_\alpha p^\alpha + S^{\alpha\beta}\nabla_\alpha\xi_\beta \right) = 0.
\end{equation}
Thus, we have demonstrated that the scalar 
\begin{equation}\label{conserved}
2\xi_\alpha p^\alpha + S^{\alpha\beta}\nabla_\alpha\xi_\beta = {\rm const}
\end{equation}
is conserved. Quite remarkably, no supplementary condition is needed. For other conserved quantities, non-linear in spin, see \cite{Ruediger:1981,Ruediger:1983}.  

\subsection{Mass parameters $M$ and $m$}

Here we study whether the mass parameters are constant or may change along body's trajectory. Recall the definitions $m = p_\alpha u^\alpha$ and $M^2 = p_\alpha p^\alpha$.

Equation (\ref{dP}) yields $u_\alpha\dot{p}^\alpha =0$. Consequently, 
\begin{equation}\label{dotm}
\dot{m} = p^\alpha\dot{u}_\alpha = \dot{u}_\alpha \frac{D}{ds}(S^{\alpha\beta}u_\beta).
\end{equation}
In the last equality we used (\ref{pmu}). As we see, $m$ is constant when the Frenkel condition (\ref{FC}) is assumed.

If we contract (\ref{dS}) with $p_\alpha\dot{p}_\beta$, we find
\begin{equation} 
\dot{p}_\beta \frac{D}{ds}\left(p_\alpha S^{\alpha\beta}\right) + mM\dot{M} = 0.
\end{equation}
Accordingly, $M$ is constant for the Tulczyjew condition (\ref{TC}).

\subsection{Velocity and momentum}

The relation between the velocity of the body and its momentum provides a very useful information for the integration of the equations of motion. Here we derive this relation explicitly without using the supplementary conditions. 

Contracting (\ref{dS}) with $p_\beta$ and making use of (\ref{dP}), we find
\begin{equation}\label{UP}
u^\alpha = {\frac {m}{M^2}}\,p^\alpha - {\frac 1{M^2}}\frac{D}{ds}\left(S^{\alpha\beta}p_\beta \right) - {\frac{1}{2M^2}}\,S^{\alpha\gamma}S^{\mu\nu}R_{\mu\nu\beta\gamma}
\,u^\beta.
\end{equation}

Now let us derive a useful identity. For an arbitrary skew-symmetric tensor $\varphi_{\alpha\beta}$ we define the dual $\varphi^{*\alpha\beta} := \frac{1}{2} \eta^{\alpha\beta\mu\nu}\varphi_{\mu\nu}$. Any two such tensors together with their duals satisfy the identity (a somewhat lengthy but straightforward proof is based on the properties of the Kronecker and Levi-Civita objects, see \cite{Synge:Schild:1949} chapter 7, e.g.)
\begin{equation}
\varphi^{\alpha\gamma}\sigma_{\beta\gamma} - \sigma^{*\alpha\gamma} \varphi^*_{\beta\gamma} \equiv \frac{ 1}{2}\,\delta^\alpha_\beta \,\varphi^{\mu\nu}\sigma_{\mu\nu}.\label{id1}
\end{equation}
Taking $\sigma = \varphi^*$ and noticing that $\varphi^{**} = - \varphi$, we obtain from (\ref{id1}) a new identity
\begin{equation}
\varphi^{\alpha\gamma}\varphi^*_{\beta\gamma} \equiv \frac{1}{4}\,\delta^\alpha_\beta\,\varphi^{\mu\nu}\varphi^*_{\mu\nu}.\label{id2}
\end{equation}
These identities underlie the following important algebraic result. Let us consider the second rank tensor 
\begin{equation}\label{Kab}
K^\alpha{}_\beta := \delta^\alpha_\beta + \varphi^{\alpha\gamma}\sigma_{\beta\gamma}.
\end{equation}
Here $\varphi^{\alpha\gamma}$ and $\sigma_{\beta\gamma}$ are the two arbitrary skew-symmetric tensors. Then the inverse of (\ref{Kab}) reads 
\begin{equation}
\left(K^{-1}\right)^\alpha{}_\beta = {\frac {\left[1 + \frac{1}{2}(\varphi\sigma)\right] \delta^\alpha_\beta - \varphi^{\alpha\gamma}\sigma_{\beta\gamma}} {1 + \frac{1}{2}\left(\varphi\sigma\right) - \frac{1}{16}\left(\varphi\varphi^*\right) \left(\sigma\sigma^*\right)}}.\label{KIab} 
\end{equation}
Here $\left(\varphi\sigma\right) = \varphi^{\mu\nu}\sigma_{\mu\nu}$ and $\left(\varphi\varphi^*\right) = \varphi^{\mu\nu}\varphi^*_{\mu\nu}$. The proof is straightforward: one should multiply $(K^{-1})^\alpha{}_\gamma K^\gamma{}_\beta$ and make use of both identities (\ref{id1}) and (\ref{id2}).

We can write (\ref{UP}) as the algebraic system 
\begin{equation}
K^\alpha{}_\beta u^\beta = \frac {m}{M^2}\,p^\alpha - \frac{1}{M^2} \frac{D}{ds}(S^{\alpha\beta}p_\beta),
\end{equation}
where $\varphi^{\alpha\beta} = S^{\alpha\beta}$ and $\sigma_{\alpha\beta} = {\frac 1{2M^2}}S^{\mu\nu}R_{\mu\nu\alpha\beta}$. Accordingly, we find $\left(\varphi\sigma\right)= {\frac 1{2M^2}}S^{\mu\nu}S^{\alpha\beta}R_{\mu\nu\alpha\beta}$, 
for example. 

Then we find that the velocity of a test body can be expressed in terms of its momentum and spin:
\begin{equation}
u^\alpha = \left(K^{-1}\right)^\alpha{}_\beta\,\hat{p}^\beta,\qquad \hat{p}^\alpha = {\frac {m}{M^2}}\,p^\alpha - \frac{1}{M^2} \frac{D}{ds}(S^{\alpha\beta}p_\beta).
\end{equation}
This result is valid for {\it any} supplementary condition. For the case of the Tulczyjew condition this relation was originally derived in \cite{Kuenzle:1972,Hojman:1975,Ehlers:Rudolph:1977}. When spin satisfies (any of) the supplementary conditions (\ref{FC}) or (\ref{TC}), we have $\left(\varphi\varphi^*\right) = S^{\alpha\beta}S^*_{\alpha\beta} = 0$, and hence
\begin{equation}
u^\alpha = \hat{p}^\alpha + {\frac {2S^{\alpha\beta}S^{\mu\nu}R_{\mu\nu\beta\gamma} \hat{p}^\gamma}{4M^2 + S^{\alpha\beta}S^{\mu\nu}R_{\mu\nu\alpha\beta}}}.\label{up}
\end{equation}

\section{De Sitter space: Killing vectors}\label{de_sitter_space_sec}

Let $\vartheta_\alpha$ be a coframe 1-form. The curvature of the de Sitter spacetime reads 
\begin{equation}\label{Rcon}
R_\alpha{}^\beta = {\frac 1{\ell^2}}\,\vartheta_\alpha\wedge\vartheta^\beta.
\end{equation}
Here $\ell$ is a real constant. The anti-de Sitter space arises with the help of the formal replacement $\ell^2 \rightarrow - \ell^2$. In components, $R_\alpha{}^\beta = {\frac 12}R_{\mu\nu\alpha}{}^\beta\vartheta^\mu \wedge\vartheta^\nu$, we have $R_{\mu\nu\alpha}{}^\beta = \frac{1}{\ell^2} \left(g_{\alpha\mu}\delta^\beta_\nu - g_{\alpha\nu}\delta^\beta_\mu\right)$. 

The de Sitter spacetime has many faces. Depending on the choice of the local coordinates, the metric can have static form, either isotropic or Schwarzschild-like, or it can be written in the cosmological form of an expanding world. 

In static isotropic coordinates, the line-element of the de Sitter spacetime reads
\begin{equation}
ds^2 = V^2dt^2 - W^2(dx^2 + dy^2 + dz^2).\label{dsiso}
\end{equation}
The functions depend only on $r^2=x^2+ y^2+ z^2$:
\begin{equation}
V = {\frac {1 - r^2/\ell^2}{1 + r^2/\ell^2}},\qquad W = {\frac {2}{1 + r^2/\ell^2}}.\label{VW}
\end{equation}

The coordinate transformation $X^a = Wx^a, a=1,2,3,$ ($x^1 = x, x^2 = y, x^3 = z$) hence 
\begin{equation}
\rho = W\,r = {\frac {2r}{1 + r^2/\ell^2}},\label{rho}
\end{equation}
brings the line element to the standard spherically symmetric form
\begin{equation}
ds^2 = \left(1 - {\frac{\rho^2}{\ell^2}}\right)dt^2 - {\frac{d\rho^2}{1 - {\frac {\rho^2}{\ell^2}}}} - \rho^2\,d\theta^2 - \rho^2\sin^2\theta\,d\phi^2.\label{dssp}
\end{equation}
Here $\rho^2=X^2+ Y^2+ Z^2$ and spherical coordinates are introduced in the usual way by $X^a = \{X =\rho\sin\theta\cos\phi,\,Y = \rho\cos\theta\cos\phi, \,Z = \rho\cos\theta\}$. The metric (\ref{dssp}) arises from the Kottler
(Schwarzschild-de Sitter) \cite{Kottler:1918,Stephani:etal:2003} metric when the mass of the central source is zero.

Another change of coordinates from the static to the cosmological frame
\begin{eqnarray}
\tilde{t} &=& t + {\frac \ell 2}\,\log\left(1 - \rho^2/\ell^2\right),\\
\tilde{X}^a &=& {\frac {e^{-t/\ell}\,X^a}{\sqrt{{1 - \rho^2/\ell^2}}}},\qquad a = 1,2,3,\label{tx}
\end{eqnarray}
brings the de Sitter metric into the form of an exponentially expanding world
\begin{equation}\label{dscos}
ds^2 = d\tilde{t}^2 - e^{2\tilde{t}/\ell}\,\delta_{ab}d\tilde{X}^ad\tilde{X}^b.
\end{equation}

The de Sitter spacetime can be viewed as a hyperboloid embedded into the flat 5-dimensional spacetime. Let ${\cal X}^A, A=0,1,2,3,4,$ be the coordinates and the line element $ds^2 = \eta_{AB}d{\cal X}^Ad{\cal X}^B$ of such a spacetime with $\eta_{AB} = {\rm diag}(1,-1,-1,-1,-1)$. The de Sitter manifold can then be embedded in it as the hyperboloid 
\begin{equation}
\left({\cal X}^0\right)^2 - \left({\cal X}^1\right)^2 - \left({\cal X}^2\right)^2 - \left({\cal X}^3\right)^2 - \left({\cal X}^4\right)^2 = - \ell^2,\label{hy}
\end{equation}
where the embedding coordinates can be chosen for example as
\begin{eqnarray}
{\cal X}^0 &=& \ell\,\sqrt{1 - {\frac {\rho^2}{\ell^2}}}\,\sinh(t/\ell),\\
{\cal X}^4 &=& \ell\,\sqrt{1 - {\frac {\rho^2}{\ell^2}}}\,\cosh(t/\ell),\\
{\cal X}^a &=& X^a,\qquad a=1,2,3.
\end{eqnarray}

\subsection{Killing vectors}

The de Sitter manifold is a maximally symmetric space, and there are 10 Killing vector fields on it. In the {\it isotropic coordinates} $(t,x^a)$, the Killing vectors read 
explicitly:
\begin{eqnarray}
\hspace{-0.8cm}\xi^{(0)} &=& \partial_t,\\
\hspace{-0.8cm}\xi^{(a)} &=& \epsilon^{abc}x_b{\frac {\partial}{\partial x^c}},\\
\hspace{-0.8cm}\eta^{(a)}_{\pm} &=& e^{\pm t/\ell}\left[{\frac WV}\frac{x^a}{\ell} {\frac{\partial}{\partial t}} \pm \left({\frac VW}\delta^{ab} + {\frac{x^ax^b}{\ell^2}}\right){\frac{\partial}{\partial x^b}}\right].
\end{eqnarray}
In the {\it static coordinates} $(t,X^a)$, they look very similar \cite{Bokhari:Quadir:1987}
\begin{eqnarray}
\hspace{-0.6cm}\xi^{(0)} &=& \partial_t,\\
\hspace{-0.6cm}\xi^{(a)} &=& \epsilon^{abc}X_b{\frac {\partial}{\partial X^c}},\\
\hspace{-0.6cm}\eta^{(a)}_{\pm} &=& {\frac {e^{\pm t/\ell}}{\sqrt{1 - \rho^2/\ell^2}}}\left[{\frac {X^a}\ell}{\frac {\partial}{\partial t}} \pm \left(1 - {\frac {\rho^2}{\ell^2}}\right){\frac {\partial}{\partial X^a}}\right].
\end{eqnarray} 
In the {\it cosmological} setting, the Killing vectors read
\begin{eqnarray}
\xi^{(0)} &=& {\frac \partial {\partial \tilde{t}}}- {\frac {\tilde{X}^a}{\ell}}{\frac \partial {\partial \tilde{X}^a}},\\
\xi^{(a)} &=& \epsilon^{abc}\tilde{X}_b{\frac {\partial}{\partial\tilde{X}^c}},\\
\eta^{(a)}_{+} &=& {\frac \partial {\partial \tilde{X}^a}},\\
\eta^{(a)}_{-} &=& {\frac {2\tilde{X}_a}{\ell}}\left({\frac \partial {\partial \tilde{t}}} - {\frac {\tilde{X}^b}{\ell}}{\frac \partial {\partial \tilde{X}^b}}\right) \nonumber \\
&&+ \left({\frac {\tilde{\rho}^2}{\ell^2}} - e^{-2\tilde{t}/\ell}\right){\frac \partial {\partial \tilde{X}^a}}.
\end{eqnarray}
Here $\tilde{\rho}^2 = \left(\tilde{X}^a\right)^2$. The expression for $\eta^{(a)}_{-}$ is rather nontrivial. 

\subsection{Conformally flat representation}

Since the Weyl tensor for the de Sitter space is trivial, the metric can be recast into a form that is conformally flat. Explicitly,
\begin{equation}\label{confds}
ds^2 = \varphi^2\eta_{ij}dx^idx^j,\qquad \eta_{ij} = {\rm diag}(+1,-1,-1,-1).
\end{equation}
The conformal factor depends only on the 4-dimensional ``radius'' $\sigma = \eta_{ij}x^ix^j$, namely,
\begin{equation}
\varphi = {\frac{1}{1 - {\frac{\sigma}{4\ell^2}}}} = {\frac{1}{1 - {\frac{\eta_{ij}x^ix^j}{4\ell^2}}}}.\label{confi}
\end{equation}
In this representation, the Killing vectors are as follows: 
\begin{eqnarray}\label{xia}
{\underset {(\alpha)}\xi} &=& \left(1+{\frac {\sigma}{4\ell^2}}\right) \partial_\alpha - {\frac {x_\alpha x^\beta}{2\ell^2}}\,\partial_\beta,\\
{\underset {[\alpha\beta]}\xi} &=& x_\alpha\partial_\beta - x_\beta\partial_\alpha.\label{xiab}
\end{eqnarray}

\section{Integrating the equations of motion in de Sitter spacetime}\label{eom_integration_sec}

In de Sitter spacetime with the curvature (\ref{Rcon}), the equation of motion (\ref{dP}) reduces to
\begin{equation}
\dot{p}^\alpha = {\frac 1 {\ell^2}}\,S^{\alpha\beta}u_\beta.\label{dPS}
\end{equation}
The complete integration of the dynamical equations depends crucially on the supplementary condition. The two most important cases are analyzed separately below. 

\subsection{Tulczyjew condition}

Assuming (\ref{TC}), we introduce the 4-vector of spin via $\check{S}^\alpha := \eta^{\alpha\beta\mu\nu}p_\beta S_{\mu\nu}$. The inverse formula yields the spin tensor in terms of the spin vector: $S^{\alpha\beta} ={\frac {1}{2M^2}} \eta^{\alpha\beta\mu\nu}p_\mu\check{S}_\nu$. By construction, we have the 
orthogonality properties
\begin{equation}
p_\alpha \check{S}^\alpha = 0, \qquad S_{\alpha\beta}\check{S}^\alpha = 0.\label{ort1}
\end{equation}
With the help of (\ref{dPS}) and (\ref{UP}) we derive further orthogonality properties 
\begin{equation}\label{ort2}
\dot{p}_\alpha\check{S}^\alpha = 0,\qquad u^\alpha\check{S}_\alpha = 0. 
\end{equation}
An immediate consequence is the (covariant) constancy of the spin vector. Indeed, we find 
\begin{eqnarray}
\dot{\check{S}}{}^\alpha &=& {\frac {1}{2M^2}}\eta^{\alpha\beta\mu\nu}\dot{p}_\beta\eta_{\mu\nu\rho\sigma} p^\rho\check{S}^\sigma \nonumber\\
&=& {\frac {1}{M^2}}\left(p^\beta\dot{p}_\beta\check{S}^\alpha - p^\alpha \dot{p}_\beta\check{S}^\beta\right) = 0.\label{dotS1}
\end{eqnarray}
Using the orthogonality properties, we find in the de Sitter space
\begin{eqnarray}
&&S^{\alpha\gamma}S^{\mu\nu}R_{\mu\nu\beta\gamma} \nonumber \\
  &&= {\frac {1}{2M^4\ell^2}}\left(p^\alpha p_\beta\check{S}^2 - \delta^\alpha_\beta \,M^2\check{S}^2 + M^2\check{S}^\alpha\check{S}_\beta\right). 
\end{eqnarray}
Consequently, 
\begin{equation}
S^{\alpha\gamma}S^{\mu\nu}R_{\mu\nu\beta\gamma}u^\beta = {\frac {\check{S}^2} {2M^2\ell^2}}\left({\frac {m}{M^2}}\,p^\alpha - u^\alpha\right). 
\end{equation}
Substituting this into (\ref{UP}), we have 
\begin{equation}
\left(1 - {\frac {\check{S}^2}{4M^4\ell^2}}\right) \left({\frac {m}{M^2}}\,p^\alpha - u^\alpha\right) = 0. 
\end{equation}
Thus, since the spin vector is spacelike, we can never have $\check{S}^2 = 4M^4\ell^2$, and hence we obtain
\begin{equation}
p^\alpha = m\,u^\alpha,\qquad m = M.\label{pamu}
\end{equation}
As a result, the equations of motion of a spinning test body in the de Sitter spacetime under the Tulczyjew condition reduce to
\begin{equation}
\dot{p}^\alpha \stackrel{(\ast\ast)}{=} 0,\qquad \dot{\check{S}}{}^\alpha \stackrel{(\ast\ast)}{=} 0,\qquad p^\alpha \stackrel{(\ast\ast)}{=} m\,u^\alpha, \label{em1}
\end{equation}
or, equivalently
\begin{equation}
\dot{u}^\alpha \stackrel{(\ast\ast)}{=} 0, \qquad \eta^{\alpha \beta \gamma \delta} u_\beta \dot{S}_{\gamma \delta} \stackrel{(\ast\ast)}{=} 0.
\end{equation}
The first equation actually means that the trajectories of the spinning bodies are the geodesics in the de Sitter space. The second equation describes the precession of the spin vector, or tensor, of a body during its motion along a geodesic curve.

\subsection{Frenkel-Pirani condition}

Let us now analyze the Frenkel case (\ref{FC}). Although the dynamic equations for this supplementary condition have a certain similarity to the above case, there are important differences. In particular, from (\ref{dPS}) it immediately follows that, like in the previous case, the momentum is covariantly constant, $\dot{p}^\alpha = 0$. 

Following the same line of reasoning, we define the 4-vector of spin by ${S}^\alpha := \eta^{\alpha\beta\mu\nu}u_\beta S_{\mu\nu}$. The inverse formula yields the spin tensor in terms of the spin vector: $S^{\alpha\beta} = {\frac{1}{2}}\eta^{\alpha\beta\mu\nu}u_\mu{S}_\nu$ (we use the normalization $u^2 = 1$). By construction, we thus have the 
orthogonality properties
\begin{equation}
u_\alpha {S}^\alpha = 0, \qquad S_{\alpha\beta}{S}^\alpha = 0.\label{ort3}
\end{equation}
Now in complete analogy with (\ref{dotS1}), directly from the definition of the spin vector, we derive that the spin vector is Fermi-Walker transported:
\begin{eqnarray}
\dot{S}{}^\alpha &=& {\frac {1}{2}}\eta^{\alpha\beta\mu\nu}\dot{u}_\beta\eta_{\mu\nu\rho\sigma} u^\rho{S}^\sigma =  - u^\alpha \dot{u}_\beta{S}^\beta.\label{dotS2}
\end{eqnarray}
We thus have the system\footnote{Here $\rho^\alpha_\beta:=\delta^\alpha_\beta - u^\alpha u_\beta$.}
\begin{equation}\label{em2}
\dot{p}^\alpha = 0,\qquad \rho^\alpha_\beta \dot{S}^\beta = 0.
\end{equation}
Although this looks formally similar to (\ref{em1}), the actual dynamics is very different. In particular, the trajectories are no longer geodesics because the momentum does not coincide with the velocity. Instead,
\begin{equation}
p^\alpha = mu^\alpha - {S}^{\alpha\beta} \dot{u}_\beta,\label{pmus}
\end{equation}
and this relation must accompany the integration of the system (\ref{em2}). The above system can be simplified even further. If we differentiate eq.\ (\ref{pmus}) and contract it with $S_\alpha$, we obtain -- with the help of (\ref{ort3}): $\dot{p}^\alpha S_\alpha =  m\dot{u}^\alpha S_\alpha - \dot{S}^{\alpha\beta} \dot{u}_\beta S_\alpha$. The last term vanishes when we use (\ref{dS}), and the left-hand side vanishes because of the covariant constancy of the momentum, cf.\ (\ref{em2}). Thus, we obtain $\dot{u}{}^\alpha S_\alpha=0$, in other words -- taking into account (\ref{dotS2}) -- the spin is also parallely transported in the Frenkel-Pirani case. We thus end up with the final system:
\begin{eqnarray}
\dot{p}^\alpha \stackrel{(\ast)}{=} 0,\quad \dot{S}^\alpha \stackrel{(\ast)}{=} 0,\quad p^\alpha \stackrel{(\ast)}{=} mu^\alpha - {S}^{\alpha\beta} \dot{u}_\beta. \label{em_frenkel_final}
\end{eqnarray}
Equivalently, one may look for solutions of the system
\begin{eqnarray}
S^{\alpha \beta} \ddot{u}_\beta - m \dot{u}^\alpha &\stackrel{(\ast)}{=}& 0, \\
\dot{S}^{\alpha \beta} + 2 u^{[\alpha} S^{\beta]\gamma} \dot{u}_\gamma &\stackrel{(\ast)}{=}&0.
\end{eqnarray}

The geodetic motion plus the parallel transport of the spin
\begin{equation}
\dot{u}^\alpha = 0,\qquad \dot{S}^\alpha = 0,\label{geoF}
\end{equation}
is a solution of (\ref{em_frenkel_final}), as one can immediately check. However, in general a spinning body does not move along a geodesic.

The force that pushes the body away from a geodesic is produced by its own spin, and the resulting motion is a classical analog of the Zitterbewegung. The key to the description of this motion is a new vector variable that can be introduced along the lines of the flat space discussion \cite{Kudryashova:Obukhov:2010}, i.e.
\begin{equation}
Q^\alpha := {\frac 1{M^2}}\,S^{\alpha\beta}p_\beta.\label{Qa}
\end{equation}
In view of the covariant constancy of the momentum, both invariants $M^2= p_\alpha p^\alpha$ and 
\begin{equation}\label{S2}
{\frac 12}S_{\alpha\beta}S^{\alpha\beta} = -\,{\frac {S_\alpha S^\alpha}{4}}
\end{equation}
are conserved along body's trajectory. As a result, the length of the new vector field is also constant
\begin{equation}
Q_\alpha Q^\alpha = {\frac {S_\alpha S^\alpha}{4M^2}}\left({\frac {m^2}{M^2}} - 1\right).\label{Q2}
\end{equation}
Being orthogonal to the velocity, $Q^\alpha$ is spacelike, and thus it is rotating thereby generating the helical motion of the body. This can be further clarified as follows. 

Contracting (\ref{dS}) with $p_\beta$, we find 
\begin{equation}
u^\alpha + \dot{Q}{}^\alpha - {\frac {mp^\alpha}{M^2}} = 0.\label{uQ}
\end{equation}
By differentiating we thus prove that the acceleration is produced by the ``$Q$-force''
\begin{equation}
\dot{u}{}^\alpha = -\,\ddot{Q}{}^\alpha.\label{ddQ}
\end{equation}

On the other hand, every vector can be expanded with respect to the natural orthogonal basis formed by the quadruple $(p^\alpha, S^\alpha, Q^\alpha,\dot{Q}{}^\alpha)$. This basis is not orthonormal since the lengths of the legs are not equal $\pm 1$, but one can straightforwardly verify that each vector is orthogonal to the three others. In particular, we can expand the acceleration with respect to this basis:
\begin{equation}
\dot{u}{}^\alpha = \alpha\,p^\alpha + \beta\,S^\alpha + \gamma\,Q^\alpha + \delta\,\dot{Q}{}^\alpha.\label{expu}
\end{equation}
Contracting this with $p_\alpha, S_\alpha, \dot{Q}{}_\alpha$, we find $\alpha = \beta = \delta = 0$ since $\dot{u}{}^\alpha p_\alpha = \dot{u}{}^\alpha S_\alpha = \dot{u}{}^\alpha\dot{Q}{}_\alpha = 0$. The only nontrivial coefficient is thus 
\begin{equation}
\gamma = {\frac {\dot{u}{}^\alpha Q_\alpha} {Q^2}}. 
\end{equation}
However, contracting (\ref{pmus}) with $p_\alpha$, we find $\dot{u}{}^\alpha Q_\alpha = 1 - m^2/M^2$, and then using (\ref{Q2}), we have $\gamma = 
-\,4M^2/S^2$. Accordingly, (\ref{expu}) reduces to
\begin{equation}
\dot{u}{}^\alpha = -\,{\frac {4M^2} {S^2}}\,Q^\alpha.\label{duQ}
\end{equation}
Comparing (\ref{ddQ}) with (\ref{duQ}), we derive the oscillator equation
\begin{equation}
\ddot{Q}{}^\alpha + \omega^2\,Q^\alpha = 0.\label{oscQ}
\end{equation}
The frequency is defined by
\begin{equation}
\omega := {\frac {2M} {\sqrt{-S_\alpha S^\alpha}}} = {\frac{M}{\sqrt{\frac{1}{2}S_{\alpha\beta}S^{\alpha\beta}}}}.\label{omega} 
\end{equation}
Furthermore, substituting $u^\alpha$ from (\ref{uQ}) into the equations of motion and the Frenkel condition, we can recast (\ref{dS}) and (\ref{FC}) into
\begin{eqnarray}
\dot{\mu}{}^{\alpha\beta} &=& 0,\label{dmu}\\
\mu^{\alpha\beta}\dot{Q}_\beta - mQ^\alpha &=& 0.\label{muQ}
\end{eqnarray}
Here we introduced another interesting object
\begin{equation}
\mu^{\alpha\beta} := S^{\alpha\beta} + p^\alpha Q^\beta - p^\beta Q^\alpha = S^{\alpha\beta} - {\frac {p^\alpha p_\gamma}{p^2}}S^{\gamma\beta} - {\frac {p^\beta p_\gamma}{p^2}}S^{\alpha\gamma},\label{mu}
\end{equation}
that is the projection of spin on the momentum. Using it, the frequency (\ref{omega}) is recast into
\begin{equation}\label{omega1}
\omega  = {\frac {m}{\sqrt{\frac{1}{2}\mu_{\alpha\beta}\mu^{\alpha\beta}}}}.
\end{equation}

Qualitatively, the dynamics of spinning bodies subject to the Frenkel condition in the de Sitter spacetime is similar to that in flat space \cite{Kudryashova:Obukhov:2010}. Everything is determined by the initial conditions. If initially (at the proper time $s=0$) spin is parallel to the momentum, i.e. $S^{\alpha\beta}p_\beta =0$ (hence $Q^\alpha =0$), then this is true on the whole trajectory that turns out to be geodesic. Otherwise, the trajectory is a geodesic curve, perturbed by the oscillatory motion of $Q^\alpha$ with the frequency (\ref{omega}) and (\ref{omega1}). 

\section{Using the integrals of motion}\label{integrals_of_motion_sec}

As a matter of fact, the de Sitter spacetime has exactly the same number of Killing vectors as the total number of the ``gravitational charges'', that is, 10. Then we can try to find the momentum $p^\mu$ and the spin $S^{\mu\nu}$ without solving differential equations by just making use of the 10 conservation laws. 

This task is most straightforwardly treated in the conformally flat representation. By substituting (\ref{xia}) and (\ref{xiab}) into (\ref{conserved}), we
have the algebraic system
\begin{eqnarray}\label{Pi}
2{\underset {(\alpha)}\xi}{\!}_\mu p^\mu + S^{\mu\nu}\nabla_\mu {\underset {(\alpha)}\xi}{\!}_\nu &=& 2\Pi_\alpha,\\ \label{Sigma}
2{\underset {[\alpha\beta]}\xi}{\!}_\mu p^\mu + S^{\mu\nu} \nabla_\mu{\underset {[\alpha\beta]}\xi}{\!}_\nu &=& 2\Sigma_{\alpha\beta}.
\end{eqnarray}
Here $\Pi_\alpha$ and $\Sigma_{\alpha\beta}= -\Sigma_{\beta\alpha}$ are the 4 + 6 = 10 constants of motion. We introduced the factors 2 for convenience. It is worthwhile to notice that due to the skew symmetry of spin, $S^{\mu\nu}\nabla_\mu\xi_\nu  = S^{\mu\nu}\partial_\mu\xi_\nu$. 

For the translational (\ref{xia}) and rotational (\ref{xiab}) Killing vectors we find
\begin{eqnarray}
{\underset {(\alpha)}\xi}{\!}_\nu = g_{\nu\lambda}{\underset {(\alpha)}\xi}{\!}^\lambda &=& \varphi^2\left[\left(1 + {\frac{\sigma}{4\ell^2}}\right) \eta_{\nu\alpha} - {\frac {x_\nu x_\alpha}{2\ell^2}}\right],\label{xiad}\\
\partial_{[\mu}{\underset {(\alpha)}\xi}{\!}_{\nu]} &=& {\frac{2\varphi^4}{\ell^2}} x_{[\mu}\hat{\eta}_{\nu] \alpha}, \label{dxiad}\\
{\underset {[\alpha\beta]}\xi}{\!}_\nu = g_{\nu\lambda}{\underset {[\alpha \beta]}\xi}{\!}^\lambda &=& \varphi^2\left(x_\alpha\eta_{\beta\nu} - x_\beta\eta_{\alpha\nu}\right),\label{xiabd}\\
\partial_{[\mu}{\underset {[\alpha\beta]}\xi}{\!}_{\nu]} &=& 2\varphi^4 \,\hat{\eta}_{\alpha[\mu}\hat{\eta}_{\nu]\beta}.\label{dxiabd}
\end{eqnarray}
Here we introduced 
\begin{equation}
\hat{\eta}_{\mu\nu} := \left(1 - {\frac {\sigma}{4\ell^2}}\right) \eta_{\mu\nu} + {\frac {x_\mu x_\nu}{2\ell^2}}.\label{heta}
\end{equation}
In addition, we define a similar object
\begin{equation}
\check{\eta}_{\mu\nu} := \left(1 + {\frac{\sigma}{4\ell^2}} \right) \eta_{\mu\nu} - {\frac{x_\mu x_\nu}{2\ell^2}}.\label{ceta}
\end{equation}
It is easy to check that up to a factor they are inverse to each other,
\begin{equation}\label{etet1}
\hat{\eta}^{\mu\lambda}\check{\eta}_{\lambda\nu} = \left(1 - {\frac {\sigma} {4\ell^2}}\right)\left(1 + {\frac{\sigma}{4\ell^2}}\right)\delta^\mu_\nu.
\end{equation} 
Other useful relations are as follows:
\begin{eqnarray}
\hat{\eta}^{\mu\lambda}\hat{\eta}_{\lambda\nu} &=& \left(1 - {\frac {\sigma} {4\ell^2}}\right)^2 \delta^\mu_\nu + {\frac {x^\mu x_\nu}{\ell^2}},\label{etet2}\\
\check{\eta}^{\mu\lambda}\check{\eta}_{\lambda\nu} &=& \left(1 + {\frac {\sigma} {4\ell^2}}\right)^2\delta^\mu_\nu - {\frac {x^\mu x_\nu}{\ell^2}}.\label{etet3}
\end{eqnarray}
The indices are everywhere raised and lowered with the help of the flat Minkowski metric $\eta_{\mu\nu}$. 

Substituting (\ref{xiad})-(\ref{dxiabd}) into (\ref{Pi}) and (\ref{Sigma}), we derive
\begin{eqnarray}
\hspace{-1cm}\varphi^2\,\check{\eta}_{\alpha\mu}p^\mu + {\frac{\varphi^4}{\ell^2}} \,\hat{\eta}_{\alpha\nu}x_\mu S^{\mu\nu} &=& \Pi_\alpha,\label{Pi1}\\ 
\hspace{-1cm}\varphi^2\left(x_\alpha\eta_{\beta\mu} - x_\beta\eta_{\alpha\mu}\right)p^\mu + \varphi^4\,\hat{\eta}_{\alpha\mu}\hat{\eta}_{\beta\nu}\,S^{\mu\nu} &=& \Sigma_{\alpha\beta}.\label{Sigma1}
\end{eqnarray}
As a first step, contracting (\ref{Pi1}) with $\hat{\eta}^{\alpha\beta}$ and using (\ref{etet1}) and (\ref{etet2}), we find
\begin{equation}
\varphi^2\,p^\mu = {\frac {1}{1 - \left({\frac {\sigma} {4\ell^2}}\right)^2}}\left({\frac {\varphi^2}{\ell^2}}\,S^{\mu\nu}x_\nu + \hat{\eta}^{\mu\nu}\,\Pi_\nu\right).\label{varp}
\end{equation}
As a result, we have
\begin{eqnarray}
&&\varphi^2\left(x_\alpha\eta_{\beta\mu} - x_\beta\eta_{\alpha\mu}\right)p^\mu = {\frac {1}{1 + {\frac {\sigma}{4\ell^2}}}}\Big[x_\alpha\Pi_\beta - x_\beta \Pi_\alpha\nonumber\\
&& + {\frac{\varphi^3}{\ell^2}}\left(x_\alpha \eta_{\beta\mu} - x_\beta \eta_{\alpha\mu} \right) x_\nu S^{\mu\nu} \Big]. \label{xpxp}
\end{eqnarray}
On the other hand, 
\begin{eqnarray}
\varphi^2 {\eta}_{\alpha\mu}{\eta}_{\beta\nu} S^{\mu\nu}  - {\frac{\varphi^3}{2\ell^2}} \left(x_\alpha \eta_{\beta\mu} -x_\beta \eta_{\alpha\mu}\right) x_\nu S^{\mu\nu} \nonumber\\
= \varphi^4\,\hat{\eta}_{\alpha\mu}\hat{\eta}_{\beta\nu}\,S^{\mu\nu}. \label{eeS}
\end{eqnarray}
With the help of (\ref{xpxp}) and (\ref{eeS}), we recast (\ref{Sigma1}) into
\begin{equation}
{\frac{\varphi^2}{\left(1+{\frac{\sigma}{4\ell^2}}\right)^2}} \, \check{\eta}_{\alpha\mu}\check{\eta}_{\beta\nu}\,S^{\mu\nu} = \Sigma_{\alpha\beta} - {\frac{1}{1+{\frac{\sigma}{4\ell^2}}}} \left(x_\alpha\Pi_\beta - x_\beta\Pi_\alpha\right).\label{Ssig}
\end{equation}
Now contracting with $\hat{\eta}^{\alpha\rho}\hat{\eta}^{\beta\sigma}$ and making use of (\ref{etet1}), we find the spin tensor explicitly
\begin{equation}\label{Smn}
S^{\mu\nu} = \hat{\eta}^{\mu\alpha}\hat{\eta}^{\nu\beta}\,\Sigma_{\alpha\beta} + \hat{\eta}^{\mu\alpha}\Pi_\alpha\,x^\nu - \hat{\eta}^{\nu\alpha}\Pi_\alpha\,x^\mu.
\end{equation}
Notice that 
\begin{equation}\label{etx}
\hat{\eta}^{\mu\alpha}x_\alpha = \left(1 + {\frac {\sigma} {4\ell^2}}\right)x^\mu,
\end{equation}
which we took into account when deriving (\ref{Smn}). Using this again, we find
\begin{equation}
S^{\mu\nu}x_\nu = {\frac 1\varphi}\left[\left(1 + {\frac {\sigma} {4\ell^2}} \right)\eta^{\mu\alpha}\Sigma_{\alpha\beta}x^\beta + (\eta^{\mu\nu}\sigma - x^\mu x^\nu)\,\Pi_\nu\right].\label{Sx}
\end{equation}
Substituting this into (\ref{varp}) we finally obtain the momentum
\begin{equation}
p^\mu = {\frac{1}{\ell^2}}\,\eta^{\mu\alpha}\Sigma_{\alpha\beta}x^\beta + \check{\eta}^{\mu\nu}\,\Pi_\nu.\label{pm}
\end{equation}

In summary, by using the 10 first integrals corresponding to the Killing vectors of the de Sitter spacetime, we are able to express the momentum and the spin as functions of the constants of motion:
\begin{eqnarray}
p^\mu &=& {\frac {1}{\ell^2}}\,\eta^{\mu\alpha}\Sigma_{\alpha\beta}x^\beta + \check{\eta}^{\mu\nu}\,\Pi_\nu,\label{Pdes}\\ 
S^{\mu\nu} &=& \hat{\eta}^{\mu\alpha}\hat{\eta}^{\nu\beta}\,\Sigma_{\alpha\beta} + \hat{\eta}^{\mu\alpha}\Pi_\alpha\,x^\nu - \hat{\eta}^{\nu\alpha}\Pi_\alpha\,x^\mu.\label{Sdes}
\end{eqnarray}
Remarkably, the dependence on the spacetime coordinates is merely polynomial. Notice that two different etas appear in the final formulas: $\check{\eta}^{\mu\nu}$ enters (\ref{Pdes}) but $\hat{\eta}^{\mu\nu}$ enters (\ref{Sdes}).

This seems to be the best what one can do without imposing the supplementary condition on spin. The dynamics of momentum and spin is completely known, given by (\ref{Pdes}) and (\ref{Sdes}), however, the trajectory of the body is still undefined. At first sight, one may think that it is possible to substitute spin and momentum into (\ref{pmu}) and solve the resulting algebraic equation for the 4-velocity $u^\alpha$. When this is done, one can find the trajectory from the velocity vector field. But this plan does not work because one can verify by direct substitution of (\ref{Pdes}) and (\ref{Sdes}) into (\ref{pmu}) that the latter is an identity.

After imposing the supplementary condition, everything reduces to a mere technical problem of integrating the first order system. For the Tulczyjew condition, such a system is obtained by substituting (\ref{Pdes}) into the left-hand side of (\ref{pamu}) and recalling that $u^\alpha = dx^\alpha/ds$ on the right-hand side. Similarly, for the Frenkel condition one needs to plug (\ref{Pdes}) into (\ref{uQ}), with an intermediate step of constructing $Q^\alpha$ by contracting (\ref{Pdes}) with (\ref{Sdes}) and substituting the result into (\ref{uQ}). In both cases, for the Frenkel and the Tulczyjew condition, the resulting first order system involves only polynomial functions of $x^\alpha$ and can be straightforwardly integrated numerically. 

\section{Discussion}\label{conclusions_sec}

We have studied the dynamics of spinning test bodies in the de Sitter spacetime. Qualitatively, the results are similar to those obtained in flat spacetime. For the Tulczyjew condition (\ref{TC}), the body moves along the geodesic curve, whereas the spin vector is parallelly transported along the trajectory. In the Frenkel case (\ref{FC}), the spin is still parallelly transported, but geodesic motion is just one special solution of the equations of motion. When the initial value of $Q^\alpha = S^{\alpha\beta}p_\beta/p^2$ is nontrivial, then the body is affected by the spin-dependent force, the acceleration $\dot{u}^\alpha$ is nontrivial, and the trajectory oscillates around a geodesic with the frequency (\ref{omega}), (\ref{omega1}). The curvature of the de Sitter space thus affects the dynamics only indirectly through the structure of the corresponding geodesics on this manifold. The introduction of the variables $\mu_{\alpha\beta}$ and $Q^\alpha$ can be qualitatively compared to the definition of the mean spin and the mean position operators in quantum mechanics \cite{Newton:Wigner:1949,Foldy:Wouthuysen:1950,Hilgevoord:Wouthuysen:1963,Hilgevoord:DeKerf:1965}.

In summary, different supplementary conditions do lead to fundamental changes on the level of the equations of motion. In particular we have explicitly shown for the de Sitter spacetime, how the solutions for the worldline change under the Tulczyjew (\ref{TC}) and the Frenkel (\ref{FC}) condition. The search for further solutions of the multipolar equations of motions is an ongoing task, in particular one should aim for a solution of the equations in more complicated spacetimes than the one covered in the present work.

As mentioned in the introduction there also remain conceptual questions to be answered in the context of different multipolar approximation schemes. In particular the interpretation of the quantities and the structure of the final set of equations of motion in such schemes should be investigated. Without going into detail at this point -- see the upcoming work \cite{Puetzfeld:2010} in this respect -- we have to stress that the system (\ref{dP})--(\ref{dS}) can be obtained in several different ways. While one can achieve formal equivalence on the level of the equations of motion, in particular up to the pole-dipole order, there are differences in the derivation as well as in the interpretation of these equations. 

Let us briefly mention two viewpoints here. Most interestingly, the equations (\ref{dP}) and (\ref{dS}) have been interpreted in a point-particle, as well as in an extended body context. In the point-particle picture it is immediately clear that the dynamics of the particles under consideration is directly influenced by the choice of the supplementary condition. The particle is thought to be localized at the worldline, and the worldline represents an immediate description of its motion through spacetime. This should eventually lead to small, but physically detectable changes in the motion of these particles. For an indirect detection one may imagine a charged particle moving in an electromagnetic field. The differences in its motion due to the different supplementary conditions -- recall the oscillatory motion in case of the Frenkel condition -- should be detectable via radiation losses. For an extended body the impact of different supplementary conditions may be less straightforward to detect. The supplementary condition can be interpreted as the choice of a suitable representative worldline of the body under consideration -- in analogy to the dynamics of bodies in Newtonian physics one may think of an extension of the concept of the center-of-mass. While it is clear that the shape of this worldline will change under different supplementary conditions, the question remains up to which level such a change impacts observable quantities. Microscopic oscillations along the representative line probably do not play any role in the description of the motion of, say, an extended star. Nevertheless one has to check for each application, if the length scales of the system under consideration justify to view the choice of the supplementary condition as something which has no direct physical impact. 

Let us close by pointing out, that further work on the foundations and the interpretation of multipolar schemes is needed. This concerns both, i.e.\ the point-particle as well as extended body, interpretations. In particular, one should always keep in mind that the multipolar schemes, which lead to equations of motion (\ref{dP}) and (\ref{dS}), are {\it approximation schemes} which -- by construction -- only capture certain features of the full theory. 

Although the present study is mainly of theoretical nature, one should keep in mind that a deeper understanding of the dynamics of spinning bodies in curved spacetime is important for many physical applications. In particular, the spin effects are well observable in astrophysical situations for the binary star systems \cite{Blandford:1995,Stella:Vietri:1998,Hotan:Bailes:Ord:2005}. Furthermore, the use of the conserved quantities (\ref{conserved}) appears to be quite useful also in situations when the number of Killing vectors is insufficient for the complete integration of the equations of motion, cf.\ for example the study of spinning bodies in the gravitational fields of
the black body type sources in \cite{Mortazavimanesh:Mohseni:2009,Mohseni:2010}.

\begin{acknowledgments}
The authors are grateful to F.W.\ Hehl (Univ.\ Cologne) for stimulating discussions and constructive criticism. DP acknowledges the support by the Deutsche Forschungsgemeinschaft (DFG) through the SFB/TR7 ``Gravitational Wave Astronomy''. YNO is grateful to the Albert-Einstein-Institute (Golm) for the
warm hospitality and support. AEI publication number 2010 - 152. 
\end{acknowledgments}

\appendix

\section{Dimensions \& Symbols}\label{app_sec}

In order to fix our notation, we provide some tables with definitions in this appendix. The dimensions of the different quantities appearing throughout the work are displayed in table \ref{tab_dimensions}. Table \ref{tab_symbols} contains a list with the most important symbols used throughout the text. Greek indices denote 4-dimensional indices and run from $\alpha = 0, \dots, 3$, the signature is (+,--,--,--). 

\begin{table}
\caption{\label{tab_dimensions}Dimensions of the quantities.}
\begin{ruledtabular}
\begin{tabular}{cl}
Dimension (SI)&Symbol\\
\hline
&\\
\hline
\multicolumn{2}{l}{{Geometrical quantities}}\\
\hline
1 & $g_{\alpha \beta}$, $\eta_{\alpha \beta}$, $\sqrt{-g}$, $\delta^\alpha_\beta$, $\vartheta^\alpha$, ${\underset{(\alpha)}{\xi_\nu}}$\\
m&$x^{\alpha}$, $dx^{\alpha}$, $s$, ${\underset{[\alpha\beta]}{\xi_{\nu}}}$ \\
m$^{-1}$& $\Gamma_{\alpha \beta}{}^\gamma$ \\
m$^{-2}$& $R_{\alpha \beta \gamma}{}^\delta$, $R_{\alpha \beta}$ \\
&\\
\hline
\multicolumn{2}{l}{{Matter quantities}}\\
\hline
1& $u^a$, $\hat{p}^\alpha$ \\
kg\,m$/$s & $m$, $M$, $p^\alpha$, $\Pi_\alpha$\\
kg\,m$^2/$s & $S^{ab}$, $S^{a}$, $\mu^{ab}$, $\Sigma_{\alpha \beta}$ \\
kg$^2$\,m$^3/$s$^2$ & $\check{S}^\alpha$\\ 
kg$/$m$^2$\,s & $T^{\alpha \beta}$\\
&\\
\hline
\multicolumn{2}{l}{{Auxiliary quantities}}\\
\hline
1 & $K^\alpha{}_\beta$, $W$, $V$, $\varphi$, $\hat{\eta}_{\alpha \beta}$, $\check{\eta}_{\alpha \beta}$\\
m & $\ell$, $\rho$, $Q^\alpha$ \\
m$^2$ & $\sigma$\\
m$^{-1}$ & $\omega$\\
m$^{-2}$ & $\gamma$\\
&\\
\hline
\multicolumn{2}{l}{{Operators}}\\
\hline
1 & $\rho^\alpha_\beta$, ``$\ast$'', $\eta^{\alpha \beta \gamma \delta}$ \\
m$^{-1}$& $\partial_i$, $\nabla_i$, $\frac{D}{ds} = $``$\dot{\phantom{a}}$'' \\
&\\
\end{tabular}
\end{ruledtabular}
\end{table}

\begin{table}
\caption{\label{tab_symbols}Directory of symbols.}
\begin{ruledtabular}
\begin{tabular}{ll}
Symbol & Explanation\\
\hline
&\\
\hline
\multicolumn{2}{l}{{Geometrical quantities}}\\
\hline
$g_{\alpha \beta}$, $\eta_{\alpha \beta}$ & Metric (general, flat)\\
$\sqrt{-g}$ & Determinant of the metric \\
$\delta^\alpha_\beta$ & Kronecker symbol \\
$\vartheta^\alpha$ & Coframe \\
${\underset{[\alpha\beta]}{\xi_{\nu}}}$, ${\underset{(\alpha)}{\xi_\nu}}$ & Killing vectors\\
$x^{\alpha}$, $s$ & Coordinates, proper time \\
$\Gamma_{\alpha \beta}{}^\gamma$ & Connection \\
$R_{\alpha \beta \gamma}{}^\delta$, $R_{\alpha \beta}$ & Curvature (tensor, 2-form) \\
&\\
\hline
\multicolumn{2}{l}{{Matter quantities}}\\
\hline
$u^a$ & Velocity \\
$m$, $M$ & Mass (Frenkel, Tulczyjew)\\
$p^\alpha$ & Generalized momentum \\
$S^{ab}$, $\mu^{ab}$ & Spin (tensor, projection) \\
$\Pi_\alpha$, $\Sigma_{\alpha \beta}$ & Constants of motion\\
$S^{a}$, $\check{S}^\alpha$ & Spin (Frenkel, Tulczyjew) \\ 
$T^{\alpha \beta}$ & Energy-momentum tensor\\
&\\
\hline
\multicolumn{2}{l}{{Auxiliary quantities}}\\
\hline
$W$, $V$, $\varphi$ & Metric coefficients (de Sitter) \\
$\sigma$ & 4d ``radius'' \\
$Q^\alpha$ & Auxiliary ``force'' variable\\
$\omega$ & Oscillation frequency\\
&\\
\hline
\multicolumn{2}{l}{{Operators}}\\
\hline
$\rho^\alpha_\beta$ & Spatial projector \\
$\eta^{\alpha \beta \gamma \delta}$ & Permutation symbol\\
``$\ast$'' & Dual \\
$\partial_i$, $\nabla_i$ & (Partial, covariant) derivative \\ 
$\frac{D}{ds} = $``$\dot{\phantom{a}}$'' & Total derivative \\
&\\
\end{tabular}
\end{ruledtabular}
\end{table}

\bibliographystyle{unsrtnat}
\bibliography{de_sitter_bibliography}

\end{document}